\documentclass[prb,twocolumn,showpacs,amsmath,amssymb]{revtex4}

\usepackage{graphicx}
\usepackage{dcolumn}
\usepackage{bm}
\usepackage{epsfig}
\usepackage{color}
\usepackage{ulem}

\begin{document}

\newcommand{\ep}{\varepsilon}
\newcommand{\udl}{\underline}
\newcommand{\ovl}{\overline}
\newcommand{\be}{\begin{eqnarray}}
\newcommand{\ee}{\end{eqnarray}}
\newcommand{\ba}{\begin{array}}
\newcommand{\ea}{\end{array}}
\newcommand{\up}{\uparrow}
\newcommand{\dn}{\downarrow}
\newcommand{\vectg}[1]{\mbox{\boldmath ${#1}$}}
\newcommand{\vect}[1]{{\bf #1}}

\title{
Spin blocker made of semiconductor double quantum well using the Rashba effect
}
\author{S. Souma$^1$}
\author{H. Mukai$^1$}
\author{M. Ogawa$^1$}
\author{A. Sawada$^2$}
\author{S. Yokota$^2$}
\author{Y. Sekine$^3$}
\author{M. Eto$^4$}
\author{T. Koga$^2$}

\affiliation{
$^1$Department of Electrical and Electronics Engineering, 
Kobe University, Nada, Kobe 657-8501, Japan
\\
$^2$Division of Electronics for Informatics, 
Graduate School of Information Science and Technology, 
Hokkaido University, Sapporo, Hokkaido 060-0814, Japan
\\
$^3$NTT Basic Research Laboratories, NTT Corporation, Atsugi, Kanagawa 243-0198, Japan
\\
$^4$
Faculty of Science and Technology, Keio University, 
Yokohama 223-8522, Japan
}

\begin{abstract}
We propose a lateral spin-{{blockade}} 
device that uses an InGaAs/InAlAs  
double quantum well (DQW), where the values of the Rashba spin-orbit 
parameter $\alpha_{\rm R}$ are opposite in sign but equal in magnitude 
between the constituent quantum wells (QW). By tuning the channel length of 
DQW and the magnitude of the externally applied in-plane magnetic field, 
one can block the transmission of one spin (e.g., spin-down) component, leading 
to a spin-polarized current. Such a spin-blocking effect, brought about by 
wave vector matching of the spin-split Fermi surfaces between the two QWs, 
paves the way for a new scheme of spin-polarized electric current generation for 
future spintronics applications based on semiconductor band engineering. 
\end{abstract}
\maketitle
\section{Introduction}

Semiconductor spintronics is a new paradigm for semiconductor electronics 
which utilizes spins of electrons in addition to charges for device 
functionalities. Primary areas for potential applications include information 
storage, computing, and quantum information. 
One of the pedagogical devices in semiconductor spintronics is the 
spin field-effect transistor (spin-FET) proposed by Datta and 
Das,\cite{datta_das} where a spin current injected from the source is 
controlled by gate via the Rashba spin-orbit 
interaction (SOI).\cite{rashba,nitta,KogaWAL}
{{Three elemental technologies are required to 
realize switching in the spin-FET.}} The first is the generation and 
injection of spins into a semiconductor. The second is the manipulation 
of spins. The third is the detection of spins. 
Among these mechanisms, the generation of spin-polarized electrons without 
conventional {{ferromagnets}} has certain merits such as efficient 
spin injection and the absence of stray magnetic 
fields. 
While there have been many proposals to fulfill this purpose,\cite{dms_Fiederling,eus} 
the uses of the SOI provide the most popular approachs. 
These include 
the intrinsic/extrinsic spin-Hall effect (SHE),\cite{Hirsch,Murakami,Kato,Wunderlich} 
spin-filtering
devices using resonnant tunneling diodes (RTD),\cite{Voskoboynikov,Koga,Ting} 
those utilizing quantum point contact
(QPC),\cite{eto,Debray,Kohda_SG,katsumoto} and topologically protected surface
current in topological insulators (TI).\cite{Kane_Mele,Bernevig} While the intrinsic
mechanisms (intrinsic SHE and TI) potentially provide an ideal (the most
efficient) source for the spin-polarized current, the actual utilization
of these are hindered by the lack of their controllability. The
extrinsic mechanisms, on the other hand, provide good controllability in
the actual devices, whereas the efficiencies of the spin current
generation are typically very low.\cite{Ando-Saitoh} 

In this report, we propose a new spin-filtering device which provides both 
excellent controllability and efficiency, based 
on the extrinsic SOI mechanism, utilizing a narrow gap semiconductor 
double quantum well (DQW) structure.\cite{Koga,matsuura_physicaE, bernardes, glazov,Ekenberg}
In this device, the combination of the Rashba SOI and the external orbital 
magnetic field enables the perfect blockade of only one spin component, thereby 
generating a spin-polarized current. 
Our DQW-based device has advantages over the QPC-based devices in a sense that 
it is compatible with the conventional lithographic top-down processes 
and that a large (spin-polarized) current can be extracted from the device. 

We organize the paper as follows. 
We introduce the concept of the 
proposed device in the next section. This is followed by a description
of tight-binding (TB) model that simulates 
the function of our device (Sec.~III), where we {{also}} show 
how the parameter values for the TB model were chosen and 
 {{that the resultant energy dispersion relation indeed agreed}} with 
that of the effective mass model.  Section~IV is devoted to the discussions of 
the spin dependent conductance and the spin-filtering effect observed 
in the proposed device. Our conclusions are given in Sec.~V. 

\section{Device concept}

\subsection{{{Device description}}}
%
\begin{figure}[t]
\vspace*{0cm}
\includegraphics[width=9cm]{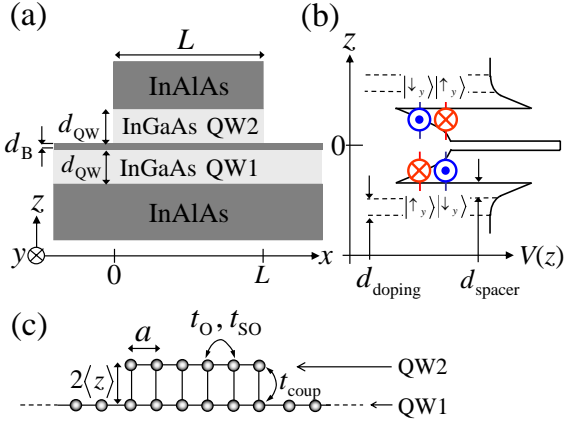}
\caption{(a) Schematic illustration of 
the proposed  spin-blocking device made of double quantum well. 
(b) Sketch of the potential profile in the DQW region with 
zero magnetic field. 
The spin-dependent energy 
eigenvalues for the given wave vector value 
$\vect{k}=(k_{\rm F},0)$ 
due to the Rashba SOI are also indicated.  
(c) The equivalent tight-binding lattice model 
to simulate 
the transport properties of the electron along $\hat{x}$.
}
\label{fig_device}
\end{figure}
\begin{figure}[h]
\vspace*{-0.5cm}
\hspace*{-0.7cm}
   \includegraphics[width=10cm]{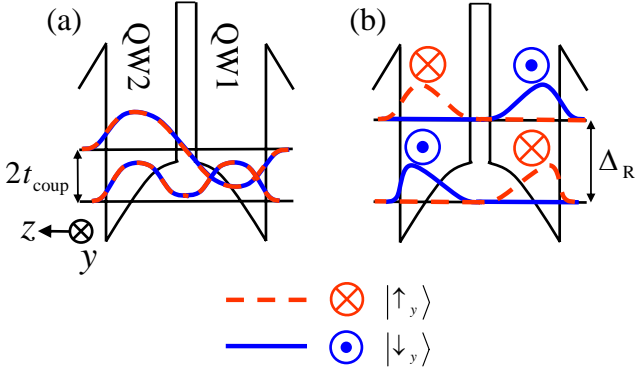}
\vspace*{-1cm}
\caption{(Color online) 
Sketches of electron wave functions along $z$-direction at $B=0$.  
(a) The case without the Rashba SOI or with 
$\vect{k}=(0,0)$ for the in-plane wave vector, 
where we obtain 
the same orbital wave functions irrespective of spins. 
(b) The case with the Rashba SOI and $\vect{k}=(k_{\rm F},0)$. 
The Rashba splitting $\Delta_{\rm R}$ is assumed to be much larger 
than the inter-well coupling $t_{\rm coup}$. 
While the spin degeneracy is preserved due to the inversion symmetry, 
the orbital wave functions are not identical between 
the different spins in each spin degenerate energy level. 
}
\label{fig_wfunc}
\end{figure}
Figure~1(a)(b) illustrates the proposed spin filtering device 
based on the DQW structure, where each QW (QW1 and QW2), by 
itself, has an asymmetric confinement potential. 
The {{locally}} averaged 
electric fields within QW1 and QW2, necessary  for the Rashba SOI for each QW, 
are denoted as $\left \langle E_z \right \rangle _{1}$ and  
$\left \langle E_z \right \rangle _{2}$, respectively.\cite{note_efield}
These electric fields satisfy the 
relation $\left \langle E_z \right \rangle _1  = - \left 
\langle E_z \right \rangle _2 > 0$ by symmetry where the $z$ 
direction is normal to the sample surface.  
Such a DQW structure can be realized using (001) InP lattice-matched 
In$_{0.53}$Ga$_{0.47}$As/In$_{0.52}$Al$_{0.48}$As material system 
as will be explained in Sec.~IIIB. 
The thicknesses of QW1 and QW2 
are both $d_{\rm QW}$, which 
are separated by a barrier layer with thickness $d_{\rm B}$. 
The value of $d_{\rm B}$ should be so chosen that QW1 and QW2 are only 
weakly coupled. 
The active part of the device, composed of DQW, has a 
length $L$ {in the $x$ direction. Non-magnetic 
electrodes are 
attached to QW1 of the DQW device at both the left and right ends, 
between which an electric current is passed 
through. 
The width of the device in the $y$ direction 
is assumed to be {{much larger than $L$}} 
so that the periodic boundary condition is applicable.

With the DQW structure described in Fig.~\ref{fig_device}(b), 
one may encounter the following 
dilemma. {{QW1 and QW2 alone are inversionally asymmetric}}, 
which {supports the presence of} the Rashba 
splitting. However, if one sees the whole DQW as a single 
quantum mechanical system, {it is inversionally symmetric},
which is against the presence of the Rashba splitting. 
Such dilemma is resolved as follows. 
If one sees QW1 and QW2 as independent quantum wells, 
the Rashba spin splitting is 
indeed induced with a finite given wave vector 
$\vect{k}=(k_{\rm F},0)$ in each QW. 
However, if one sees the DQW as a single quantum 
mechanical system, what we interpreted as the 
spin splitting of a single quantum well above 
is now viewed 
as a spin degenerate subband splitting derived from a single DQW 
(see Figs.~\ref{fig_wfunc} and \ref{fsmatching01}).

\begin{figure}[h]
\hspace*{-1cm}
\includegraphics[width=18cm]{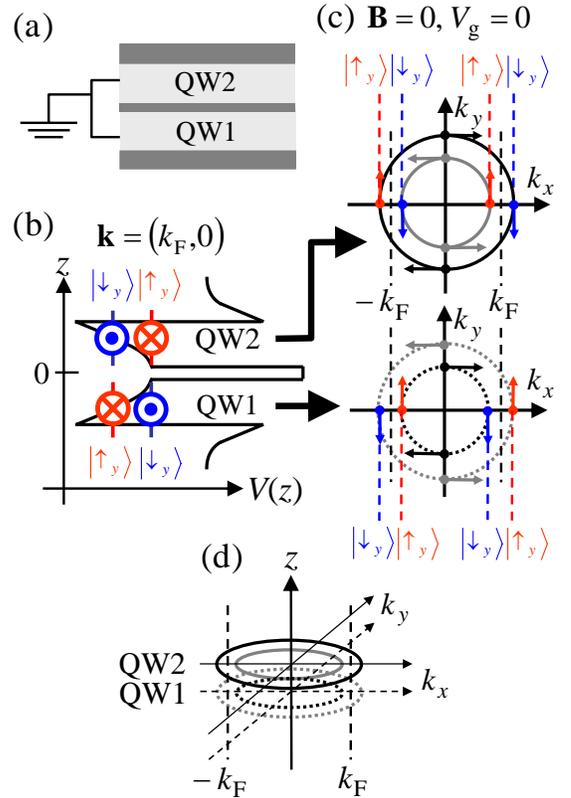}
\vspace*{-1cm}
\caption{
(Color online) 
(a) Illustration of unbiased DQW system without magnetic field. 
(b) A pair of spin degenerate eigenenergies for $\vect{k}=(k_{\rm F},0)$ in the  DQW. 
(c) Spin-dependent Fermi circles for independent 
QW1 and QW2, where indicated by the arrows on the circles are the spin orientations. 
(d) 3D illustrations of the overlapped Fermi circles of the DQW with the given 
condition, where the vertical axis is $z$ in the real space. 
}
\label{fsmatching01}
\end{figure}

\subsection{Formation of the bonding and antibonding wave functions 
for a selected spin by the in-plane magnetic field at the 
Fermi circle points $(\pm k_{\rm F},0)$ 
}
\begin{figure}[h]
\hspace*{-1cm}
\includegraphics[width=18cm]{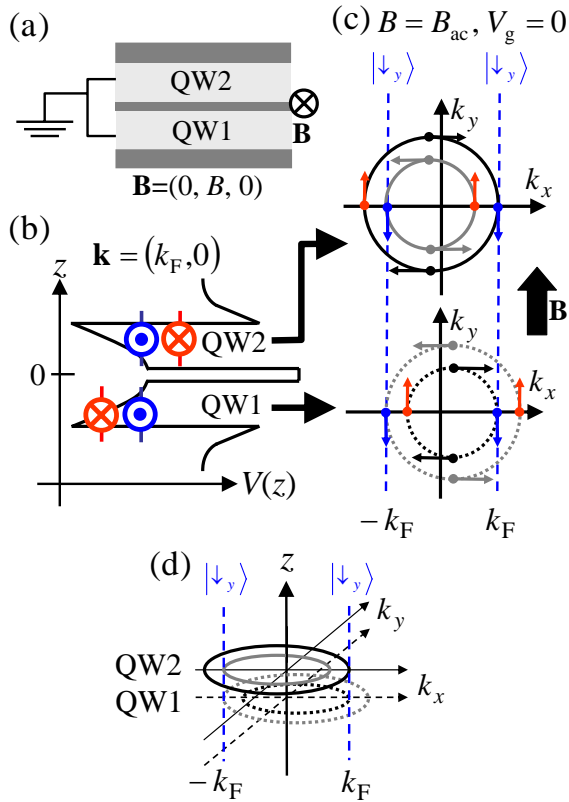}
\vspace*{-1cm}
\caption{(Color online)
(a) Illustration of unbiased DQW in 
the presence of the magnetic field 
${\bf B}=(0,B,0)$. 
(b) Spin-dependent eigenenergies in the DQW 
with $\vect{k}=(k_{\rm F},0)$ 
and $B=B_{\rm ac}$ (see text for $B_{\rm ac}$). 
(c) Spin-dependent Fermi circles for independent 
QW1 and QW2 at $B=B_{\rm ac}$, where 
we set $z=0$ at the middle of the center barrier layer. 
(d) 3D illustration of the overlapped Fermi circles of the DQW with the given condition. 
}
\label{fig_fermicircle02}
\end{figure}
Let us consider the effect of applying an in-plane magnetic 
field (${\bf B}\parallel
\hat{y}$) to the DQW, which provides a key element to the 
proposed spin-blocking mechanism. Setting 
$\vect{p}\rightarrow\vect{p}+e\vect{A}$ in the {{Hamiltonian}} {{using 
the}} vector potential $\vect{A}=(Bz,0,0)$, {{which}} leads to 
$\vect{B}=\nabla\times\vect{A}=(0,B,0)$, the Hamiltonian 
of an electron in the DQW ($H=H_0+H_{\rm R}+H_{\rm Z}$) {{reads}} 
\begin{eqnarray}
\begin{array}{c}
\displaystyle{
H_0=\frac{\hbar^2}{2m^*_{||}}\left[
\left(k_x+eBz/\hbar\right)^2
+k_y^2
\right]}
\\
\hspace{3cm}
\displaystyle{
-\frac{d}{dz}\frac{\hbar^2}{2m^*_z(z)}
\frac{d}{dz}+V(z)},
\\\hspace{-7cm}
\\
H_{\rm R}=
\alpha(z)\left[
k_y\sigma_x-\left(k_x+eBz/\hbar \right)\sigma_y
\right]
\\\hspace{-7cm}{\rm and}
\\
H_{\rm Z}=-\frac{1}{2}g^*\mu_{\rm B}B\sigma_y, 
\end{array}
\label{eq:hamiltonian_bfield}
\end{eqnarray}
where $H_0$, $H_{\rm R}$ and $H_{\rm Z}$ are the 
unperturbed Hamiltonian, Rashba and Zeeman 
{{Hamiltonians}}, respectively, 
$\mu_{\rm B}=e\hbar/2m_e$ is the Bohr magneton 
({{$m_e$ being the 
free electron mass}}), $g^*$ is the effective 
$g$-factor value, $m^*_{||}$ and $m^*_z(z)$ 
are the in-plane and out-of-plane effective masses, 
respectively, the latter incorporating different effective mass 
values for the well and barrier materials. 
$\vect{k}=(k_x,k_y)$ is the in-plane 
wave vector of an electron.\cite{note0} 
$e$ is the elementary charge. 
$\hbar$ is Planck's constant divided by 2$\pi$. 
$\sigma_x$ and $\sigma_y$ are the 
Pauli spin matrices. For the In$_{0.53}$Ga$_{0.47}$As/In$_{0.52}$Al$_{0.48}$As 
material system, 
we can safely ignore $H_{\rm Z}$ and $eBz/\hbar$ relative to $H_{\rm R}$ 
and $k_x$, respectively, 
and so we do hereafter.\cite{note1} 

Using the material-dependent intrinsic constant $a_{\rm SO}$ 
for the Rashba SOI,\cite{faniel} $\alpha (z)$ can be written as 
{$\left ( a_{\rm SO}/{{e}} \right ) 
\left ( \partial V(z)/\partial z \right )$}, 
where the contributions from the band 
discontinuities at the well-barrier interfaces
are readily included in this formulation 
{{and}} should be excluded in taking 
the derivative $\partial V(z)/\partial z$.
We also define the parameter value $\alpha_{\rm R}$ by {{
$\left \langle \Phi_{\rm QW1} \right | \alpha(z) 
\left | \Phi_{\rm QW1}\right \rangle = a_{\rm SO} \left 
\langle E_z \right \rangle_1$ or 
equivalently $-\left \langle \Phi_{\rm QW2} \right | \alpha(z) 
\left | \Phi_{\rm QW2}\right \rangle = - a_{\rm SO}\left \langle E_z 
\right \rangle_2$}}, 
where $\left|\Phi_{\rm QW1(QW2)}\right>$ is the 
energy eigenstate of the independent QW1 (QW2) in the 
confining direction.

In Eq.~(\ref{eq:hamiltonian_bfield}), we see that applying a magnetic 
field ($\vect{B}\parallel\hat{y}$) has the effect of shifting the 
Fermi circle in the $k_x$ direction by the magnitude 
$-eBz/\hbar$. Setting the origin of $z$ at the 
middle of the barrier layer, the Fermi circles for QW1 and 
QW2 are shifted oppositely along $k_x$ axis as shown in 
Fig.~\ref{fig_fermicircle02}. 
We then claim that the spin blocking effect is maximized 
when the magnitude of the Fermi circle shift becomes equal to the 
Rashba wave 
number $k_\alpha$, where $k_\alpha\equiv m^*_{||}\alpha_{\rm R}/\hbar^2$.
We call the corresponding magnetic field as the ``anticrossing'' magnetic 
field $B_{\rm ac}\equiv \hbar k_\alpha/e \left \langle z \right \rangle $, 
where $\left \langle z  \right \rangle\equiv\left \langle 
\Phi_{\rm QW2} |  z |\Phi_{\rm QW2} \right \rangle = -\left \langle 
\Phi_{\rm QW1} |  z |\Phi_{\rm QW1} \right \rangle > 0 $ (see Fig.~\ref{fig_fermicircle02}).

With this condition ($B=B_{\rm ac}$), defining the spin-up 
direction in $\hat{y}$ and neglecting the interaction between QW1 and 
QW2, the front and back edges of the Fermi circles for spin-down 
electrons in the current direction ($\parallel\hat{x}$) are at 
$\vect{k}=(k_{\rm F},0)$ and 
$(-k_{\rm F},0)$, respectively, for both QW1 and QW2.  
For spin-up electrons, those of QW1 and QW2 are at 
$(\pm k_{\rm F}+2k_\alpha,0)$  and $(\pm k_{\rm F}-2k_\alpha,0)$, 
respectively, as shown in Fig.~\ref{fig_fermicircle02}(c). 
If the interaction 
between QW1 and QW2 is turned on, only the spin-down electrons form the 
bonding- and antibonding-like wave functions 
around the Fermi circle points 
$(\pm k_{\rm F},0)$ in the presence of  $B=B_{\rm ac}$.

\subsection{Spin-blockade mechanism explained by 
the spin-selective flying qubit model}
%
\begin{figure}[h]
\vspace*{-1cm}
\hspace*{-0.5cm}
\includegraphics[width=10cm]{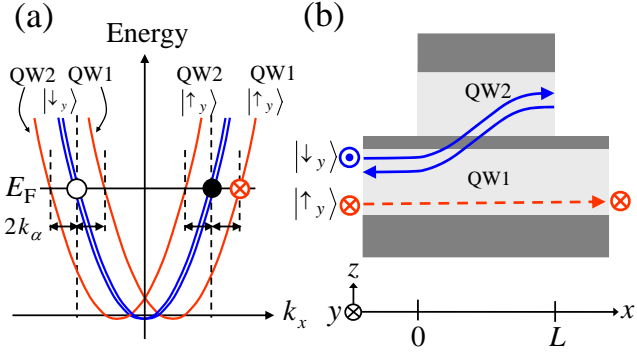}
\vspace*{-1cm}
\caption{(Color online) 
(a) Energy dispersion relation of the proposed DQW when 
the in-plane magnetic field $\vect{B}=(0,B_{\rm ac},0)$ is applied. 
(b) Spin dependent trajectories of an electron which is injected in QW1 from left, 
where the Fermi wave number matching condition is satisfied 
with the in-plane magnetic field $\vect{B}=(0,B_{\rm ac},0)$. 
}
\label{fig_trajectory_bfield}
\end{figure}
Let us consider the Fermi circle points of the DQW 
on the $k_x$ axis in the presence of $\vect{B}=\left ( 
0,B_{\rm ac},0 \right )$.  The $k_x$ values of these points are 
$-\left ( k_{\rm F}+2k_\alpha \right )$ 
(spin-up), $-k_{\rm F}$ (spin-down, doubly degenerate), 
$-\left ( k_{\rm F}-2k_\alpha\right )$ (spin-up), 
$k_{\rm F}-2k_\alpha$ (spin-up), 
$k_{\rm F}$ (spin-down, doubly degenerate), and 
$k_{\rm F}+2k_\alpha$ (spin-up), if the inter-well coupling is 
neglected. 
The degeneracies at $\pm k_{\rm F}$ are 
lifted if the inter-well coupling is turned on, i.e., 
$k_{\rm F}\rightarrow k_{\rm F}\pm k_{\rm coup}$ with 
$k_{\rm coup}=t_{\rm coup}(dE(k_x)/dk_x)^{-1}$, 
where $t_{\rm coup}$ and $E(k_x)$ are a half of the subband 
splitting energy and the energy dispersion relation of the unperturbed 
hamiltonian $H_0$ with $B=0$, 
respectively. 
We note that the states with 
$\pm k_{\rm F}-2k_\alpha$ 
have a wave function along $\hat{z}$ nearly 
equal to $\left |\Phi_{\rm QW2}\right \rangle
\otimes \left | \up_y \right 
\rangle$, while that
for $\pm  k_{\rm F}+2k_\alpha$ 
is $\left |\Phi_{\rm QW1}\right \rangle \otimes \left | \up_y\right 
\rangle $. 
The Fermi circle points $\left (\pm \left ( k_{\rm F}+
k_{\rm coup} \right ),0\right )$  and  
$\left (\pm \left ( k_{\rm F}-k_{\rm coup} 
\right ),0\right )$, 
on the other hand, form bonding- and antibonding-like 
wave functions between QW1 and QW2, respectively.  
The superposition of these wave functions causes 
precessional motion of electron between QW1 and QW2 
as we see below.\cite{bilayer}

Letting $\left |\Phi_{\rm QW1}^{\dn_y}\right \rangle 
=\left |\Phi_{\rm QW1}\right \rangle \otimes \left | \dn_y \right 
\rangle ={}^t\hspace{-0.1cm}\left(1,0\right )
\otimes \left | \dn_y \right \rangle$ and 
$\left |\Phi_{\rm QW2}^{\dn_y}\right \rangle
=\left |\Phi_{\rm QW2}\right \rangle \otimes \left | \dn_y \right 
\rangle ={}^t\hspace{-0.1cm}\left(0,1\right)\otimes 
\left |\dn_y \right \rangle $ 
be the spin-down wave functions confined in QW1 and QW2, 
respectively, the bonding- and 
antibonding-like wave functions, 
which are the spin-down energy eigenstates at 
$\left ( \pm \left ( k_{\rm F} + k_{\rm coup} 
\right ),0\right )$ 
and at 
$\left ( \pm \left ( k_{\rm F} -  k_{\rm coup}\right 
),0\right )$ are, respectively, 
\begin{eqnarray}
\left|\Phi_b^{\dn_y}\right>
=
\frac{1}{\sqrt{2}}
\left(
\begin{array}{c}
1 \\1 
\end{array}
\right)\otimes \left|\dn_y\right>
=
\frac{1}{\sqrt{2}}
\left(
\left|\Phi_{\rm QW1}^{\dn_y}\right>
+
\left|\Phi_{\rm QW2}^{\dn_y}\right>
\right)\label{eq:bonding}
\end{eqnarray}
and
\begin{eqnarray}
\left|\Phi_a^{\dn_y}\right>
=
\frac{1}{\sqrt{2}}
\left(
\begin{array}{c}
1 \\ -1 
\end{array}
\right)\otimes \left|\dn_y\right>
=
\frac{1}{\sqrt{2}}
\left(
\left|\Phi_{\rm QW1}^{\dn_y}\right>
-
\left|\Phi_{\rm QW2}^{\dn_y}\right>
\right).
\nonumber \\
\label{eq:antibonding}
\end{eqnarray}
Next, we consider how {{the wave 
function  $e^{ik_{\rm F}x}\left |\Phi_{\rm 
QW1}^{\dn_y}\right >$ in the left lead of the device propagates through 
the active part of the DQW device ($0\leq x\leq L$).}} 
{{At $x=0$, 
$\left |\Phi_{\rm QW1}^{\dn_y}\right \rangle 
= \frac{1}{\sqrt{2}}\left (\left | 
\Phi_b^{\dn_y}\right> +\left|\Phi_a^{\dn_y}\right> \right )$ }}
from Eqs.~(\ref{eq:bonding}) and (\ref{eq:antibonding}), 
where the plane wave parts for $\left |
\Phi_b^{\dn_y}\right>$ and $\left|\Phi_a^{\dn_y}\right>$ are
$e^{i\left (k_{\rm F}+ k_{\rm coup}\right )x}$
and $e^{i\left (k_{\rm F}- k_{\rm coup}\right )x}$, 
respectively, {{at the Fermi energy $E_{\rm F}$}}. 
Thus, the wave function 
$e^{ik_{\rm F}x}\left |\Phi_{\rm QW1}^{\dn_y}\right >$ in the 
left lead ($x < 0$) connects to a superpositioned state 
$\left|\Psi_{\rm inj}^{\dn_y}\left ( x \right ) \right \rangle 
\equiv \frac{e^{ik_{\rm F}x}}{\sqrt{2}}\left 
(e^{ik_{\rm coup}x} \left |\Phi_b^{\dn_y}\right >
+e^{-ik_{\rm coup}x}\left|\Phi_a^{\dn_y}\right> \right )
=e^{ik_{\rm F}x}\left \{ {\rm cos}\left ( k_{\rm coup}x 
\right ) \left|\Phi_{\rm QW1}^{\dn_y}\right >+i{\rm sin}
\left ( k_{\rm coup}x \right ) \left |\Phi_{\rm QW2}^{\dn_y}
\right>\right \}$ in the DQW ($x\geq 0$) 
[the point indicated with {\large ${\bullet}$} in Fig.~\ref{fig_trajectory_bfield}(a)]. 
This wave function portrays the precessional 
motion of an spin-down electron between QW1 and QW2 within the DQW, 
where the condition $L=L_n\equiv
\left ( n-\frac{1}{2} \right ) \pi /k_{\rm coup}$, 
$n$ being an integer, 
makes an electron be 
backscattered at the end of the DQW within QW2. 
The backscattered wave function $e^{-ik_{\rm F}\left ( x-L 
\right ) }\left |\Phi_{\rm QW2}^{\dn_y} \right >$ at $x=L$ 
now connects to 
$\frac{e^{-ik_{\rm F}x^{'}}}{\sqrt{2}}$
$\left (e^{-ik_{\rm coup}x^{'}
}
\left | \Phi_b^{\dn_y}\right>-e^{ik_{\rm coup}x^{'}
}
\left | \Phi_a^{\dn_y}\right> \right )=e^{-ik_{\rm F}x^{'}}
\left \{ 
-i{\rm sin}\left ( k_{\rm coup}x^{'}
\right )
\left|\Phi_{\rm QW1}^{\dn_y}\right >+{\rm cos}\left ( 
k_{\rm coup}
x^{'} \right ) \left | \Phi_{\rm QW2}^{\dn_y}\right>\right 
\}$ in the DQW [the point indicated with {\large $\circ$} in Fig.~\ref{fig_trajectory_bfield}(a)], 
where $x^{'}=x-L$, which propagates back to 
QW1 at $x=0$ following the trajectory 
of $\left|\Psi_{\rm inj}^{\dn_y}\left ( x \right ) \right \rangle$ backward 
as shown by the arrowed curves marked with $\odot$ in Fig.~\ref{fig_trajectory_bfield}(b).

Electrons with up-spin, on the other hand, injected from 
the left lead to QW1, are transmitted straight 
to the right lead, i.e., 
$\left|\Psi_{\rm inj}^{\up_y}\left ( x \right ) \right \rangle 
\equiv  e^{i \left ( k_{\rm F} + 2k_\alpha \right )x} 
\left |\Phi_{\rm QW1}^{\up_y}\right \rangle $ as shown by the 
arrowed broken line 
in Fig.~\ref{fig_trajectory_bfield}(b). 

The probabilities of finding spin-down and spin-up electrons 
in QW1 at the position $x=L$ upon injecting the same 
spin in QW1 at position $x=0$ are given naively as  
\begin{equation}
T_{\up(\rm QW1)}=
\left|
\left<\Phi_{\rm QW1}^{\up_y}\right|
\left.\Psi_{\rm inj}^{\up_y}(L)\right>\right|^2=1,
\label{eq:Transmission}
\end{equation}
and
\begin{equation}
T_{\dn(\rm QW1)}=
\left|\left<\Phi_{\rm QW1}^{\dn_y}\right|
\left.\Psi_{\rm inj}^{\dn_y}(L)\right>\right|^2
=
\cos^{2}\left ( k_{\rm coup}L \right ) 
\label{eq:Transmission_up}
\end{equation}
respectively. 
This means that the transmission 
probability of spin-down electron oscillates with the 
{{device}}
length $L$, while that of spin-up electron is not (always unity).

\subsection{Failure of the spin-orbit blockade by the 
Fermi circle matching with gate}

\begin{figure}[h]
\hspace*{-1cm}
\includegraphics[width=18cm]{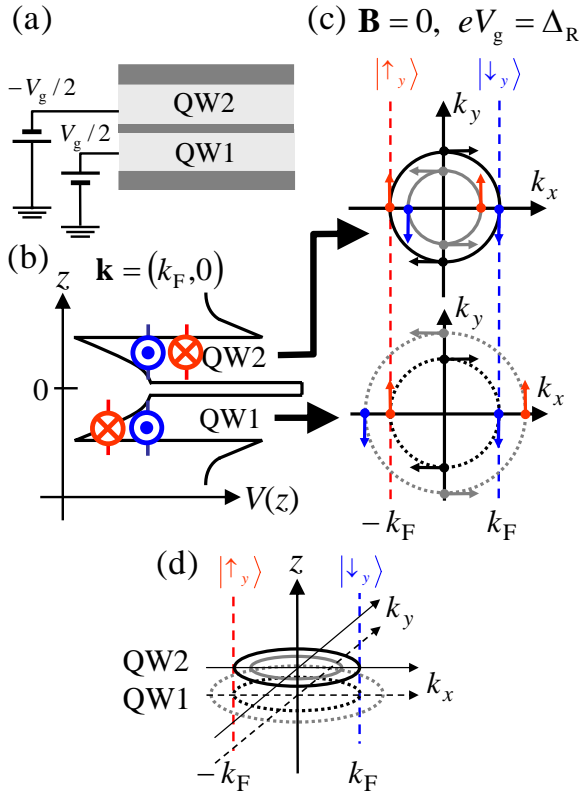}
\vspace*{-1cm}
\caption{(Color online) 
(a) Equivalent circuit model for the gated DQW without the applied magnetic field. 
(b) Spin-dependent eigenenergies in the DQW with 
$\vect{k}=(k_{\rm F},0)$ and $V_{\rm g}=\Delta_{\rm R}/e$ [see (a) for $V_{\rm g}$].  
(c)  Spin-dependent Fermi circle for independent QW1 and QW2 with 
$\vect{B}=0$ and $V_{\rm g}=\Delta_{\rm R}/e$. 
(d) 
3D illustration of the overlapped Fermi circles of the DQW with the given condition.
}
\label{fig_fermicircle03}
\end{figure}

\begin{figure}[h]
\vspace*{-0.5cm}
\hspace*{-0.5cm}
\includegraphics[width=10cm]{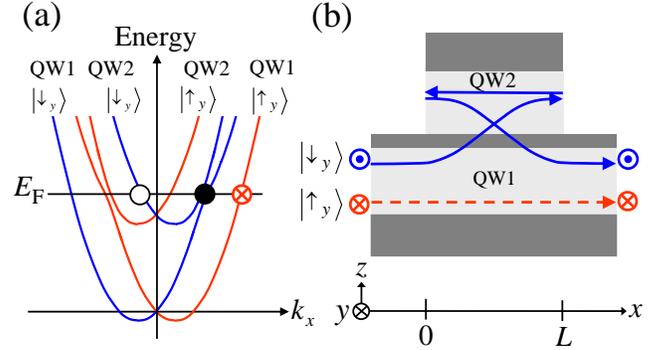}
\vspace*{-1cm}
\caption{(Color online) 
(a) 
Energy dispersion relation of the gated DQW with $V_{\rm g}=\Delta_{\rm R}/e$ 
[see Fig.~\ref{fig_fermicircle03}(a) for $V_{\rm g}$]. 
(b) 
Spin-dependent trajectories of an electrons which are injected in QW1 from left, 
where the Fermi wave number matching condition is satisfied 
by gate ($V_{\rm g}=\Delta_{\rm R}/e$).  }
\label{fig_trajectory_gate}
\end{figure}

One may think that the same spin-blocking mechanism should be 
equally effective even if the size of the 
spin-dependent Fermi circles are controlled by the surface or back gate voltages, 
not  by the in-plane magnetic field, but it is not the case as follows.  
Two Fermi 
circles of a selected spin can be made overlapped 
{{if the potential energies 
of QW1 and QW2 are shifted}} by a half of the Rashba splitting, 
i.e., $-\Delta_{\rm R}/2$ and 
$\Delta_{\rm R}/2$, respectively, or vice versa 
[see Fig.~\ref{fig_wfunc}(b) for $\Delta_{\rm R}$].   
We note that the change in the value 
of $\alpha_{\rm R}$ by this gating is almost negligible 
in the InGaAs/InAlAs DQW 
because the gate voltage required to shift the potential energies 
by $\pm\Delta_{\rm R}/2$ is very small.  
It turned out, however, 
that the proposed spin-blockade 
mechanism can not be validated  by gating 
at least within the scheme of the simple one-dimensional 
model presented here.\cite{gate,2Dmodel}

For example, 
with $-\Delta_{\rm R}/2$ and $\Delta_{\rm R}/2$ for the shift of 
the potential energies in QW1 and QW2, respectively, and with 
$L=L_n$, the injected wave function $e^{ik_{\rm F}x}\left 
|\Phi_{\rm QW1}^{\dn_y} \right \rangle$ at $x=0$ connects 
to $\left|\Psi_{\rm inj}^{\dn_y}\left ( x \right ) \right 
\rangle$, which experiences a precessional 
motion between QW1 and QW2 as in the case 
of the application of the in-plane magnetic field $B_{\rm ac}$ 
[the point indicated with {\Large $\bullet$} in Fig.~\ref{fig_trajectory_gate}(a)]. 
However, the wave function 
$\left|\Psi_{\rm inj}^{\dn_y}\left ( x \right ) \right \rangle$ 
at $x=L$ now connects to  
$e^{-i\left ( 
k_{\rm F}- 2k_{\alpha} \right ) \left ( x-L \right ) }
\left | \Phi_{\rm QW2}^{\dn_y}\right \rangle$ after 
backscattering in QW2 [{\Large $\circ$} in Fig.~\ref{fig_trajectory_gate}(a)], 
which propagates back to $x=0$ through 
QW2, where it gets reflected again.
Then, this wave function again experiences a 
precessional motion from  
QW2 to QW1 and reach $x=L$ in QW1, where the electron is ejected to the right lead 
[curves marked with $\odot$ in Fig.~\ref{fig_trajectory_gate}(b)]. 
On the other hand, for a spin-up electron injected at $x=0$,  
$e^{i\left ( k_{\rm F}+2k_\alpha \right ) x }\left 
| \Phi_{\rm QW1}^{\up_y} \right 
\rangle$ simply propagates through QW1 all the 
way to the right lead 
[broken line  in Fig.~\ref{fig_trajectory_gate}(b)]. 
Therefore, both $e^{i  k_{\rm F}  x  } \left | 
\Phi_{\rm QW1}^{\dn_y}\right \rangle$ and
$e^{i\left ( k_{\rm F}+2k_\alpha \right ) 
x }\left | \Phi_{\rm QW1}^{\up_y} \right 
\rangle$ injected from the left lead of the device are 
eventually ejected to the right lead. 
This illustrates the failing mechanism of the spin-orbit 
blockade effect by gating instead of applying 
the in-plane magnetic field within the current simple one-dimensional model.\cite{2Dmodel}

\section{Tight-binding model description of the proposed device} 

\subsection{Tight-binding model}
Keeping in mind that the proposed spin-filtering 
device has the translational symmetry along the 
$y$-direction and the front edge of the Fermi circle 
in $\hat{k}_x$ carries the most of the electric current, 
we model the device by a tight-binding (TB) 
Hamiltonian\cite{nikolic} consisting of a 
coupled one-dimensional chain as shown in Fig.~1(c), 
\begin{equation}
\hat{H}_{k_y}=\sum_{\vect{m},{\sigma},\sigma'} 
\epsilon_{\vect{m}k_y}^{\sigma\sigma'}
\hat{c}^{\dagger}_{\vect{m} {\sigma}}
\hat{c}_{\vect{m}{\sigma'}}
+
\sum_{\vect{m},\vect{m}',{\sigma},{\sigma}'} 
\hat{c}^{\dagger}_{\vect{m}{\sigma}}  
t^{\sigma\sigma'}_{\vect{m}\vect{m}'}
\hat{c}_{\vect{m}'{\sigma}'}, 
\label{eq:Hamiltonian2}
\end{equation}
where 
\begin{eqnarray}
\epsilon_{\vect{m}k_y}^{\sigma\sigma'}
&=&
\left[\frac{\hbar^2 k_y^2}{2m^*_{||}}+2t_{\rm O}
\right]
\delta_{\sigma\sigma'}+
2at_{\rm SO}(m_z)k_y\left[\sigma_x\right]_{\sigma\sigma'}
\label{eq:onsite}
\end{eqnarray}
\noindent and 
\begin{eqnarray}
t^{\sigma\sigma'}_{\vect{m}\vect{m}'}
=
\left\{
\begin{array}{ll}
\left[-t_{\rm O}\delta_{\sigma\sigma'}\mp it_{\rm SO}(m_z)\left[\sigma_y\right]_{\sigma\sigma'}\right]
\\ 
\hspace{0cm}\times \exp (\mp i\phi(m_z)) & (\vect{m}=\vect{m}'\pm \vect{e}_{x})
 \\ 
-t_{\rm{coup}}\delta_{\sigma\sigma'} & (\vect{m}=\vect{m}'\pm \vect{e}_{z})\\
0 & ({\rm otherwise}). 
\end{array}
\right.
\label{eq:hopping}
\end{eqnarray}
Here, $\hat{c}_{\vect{m}{\sigma}}$ 
($\hat{c}^{\dagger}_{\vect{m} {\sigma}}$) 
is the annihilation (creation) operator of an electron at 
site $\vect{m}=(m_x,m_z)$ with spin $\sigma$ 
($=\up,\dn$), where spin basis can be chosen in an arbitrary 
direction and $m_x$ and $m_z$ are the lattice site indices along the 
transport and out-of-plane directions, respectively. 
For $m_z$ (=1 or 2), 1 and 2 represent QW1 and QW2, 
respectively. $t_{\rm O}\equiv \hbar^2/2m^*_{||}a^2$ is the orbital 
hopping parameter where $a$ is the lattice spacing along 
the transport direction 
{{($a=1$ nm in the present work)}}. 
$t_{\rm{SO}}(m_z)=\left ( m_z -\frac{3}{2} \right ) 
\alpha_{\rm R}/a$ is the Rashba SOI hopping parameter along 
the $x$-direction within the $m_z$th QW.  The tunneling 
between QW1 and QW2 is characterized by the inter-well 
coupling parameter $t_{\rm coup}$, which is a half of the 
subband splitting for the unperturbed Hamiltonian $H_0$ in 
the absence of $B$ [see Eq.~(1) and Fig.~\ref{fig_wfunc}(a)].  
The effect of the applied magnetic field is incorporated as a form of 
Peierls phase factor 
\begin{eqnarray}
\phi(m_z)
&=&2 \pi  \frac{\Phi}{\Phi_{0}} 
\Bigl( m_z-\frac{3}{2}\Bigr),
\label{eq:Peierls}
\end{eqnarray}
where $\Phi_0=h/e$ and $\Phi=2Ba\left \langle z
\right \rangle$ are the magnetic flux quantum 
and the magnetic flux threading through the TB unit cell, 
respectively. 
The left and right leads are also described by the TB Hamiltonian 
Eq.~(\ref{eq:Hamiltonian2}) but the lattice site along the 
out-of-plane direction is restricted to $m_z=1$ (QW1), 
and the SOI coupling and magnetic field are 
{{set to}} 
zero. 
While our TB Hamiltonian [Eq.~(\ref{eq:Hamiltonian2})] together with 
Eq.~(\ref{eq:conductance}) below allows us to 
investigate on deterioration of the spin coherence by finite 
$k_y$ component in the wave vector as following, 
our investigations in this article  
are restricted to the case with $k_y=0$ to illustrate 
our fundamental idea of spin-blocking. 
First, the condition of spin-blocking for
$k_y=0$ (e.g., $B=B_{\rm ac}$ in Fig.~\ref{bdependence}) 
is different from 
that for $k_y \ne 0$ and $k_x = \pm \sqrt{k_{\rm F}^2-k_y^2}$.
Second, the first term in $H_{\rm R}$ in Eq.~(\ref{eq:hamiltonian_bfield}) 
[or the second term in Eq.~(\ref{eq:onsite})] results in the
precession of up- and down-spins in the $y$ direction.
The detailed evaluation of such anti-spin-blocking effects is
being in progress.
It should be possible to design an efficient device
whose width $W$ is smaller than its length $L$ to reduce the
anti-spin-blocking effects while $W$ is large enough to
carry a large current.

Once the Hamiltonian [Eq.~(\ref{eq:Hamiltonian2})] is set, 
the spin-dependent electric current in the right 
lead is given as $I_{\up,\dn}=G_{\up,\dn} V_{\rm SD}$ in the 
linear response regime, where $V_{\rm SD}$ is the 
{{spin-independent}} source-drain 
bias voltage. The zero-bias spin-dependent conductance $G_{\up,\dn}$ 
associated with the right lead 
is expressed as follows in the Landauer-B\"uttiker formalism.\cite{landauer}
\begin{equation}
G_{\sigma}=\frac{e^{2}}{h} \sum_{k_y}\sum_{\sigma'=\up,\dn}\int dE 
T_{\sigma\sigma',k_y}(E)\biggl\{ -\frac{df(E-E_{\rm F})}{dE}\biggr\},
\label{eq:conductance}
\end{equation}
where $T_{\sigma\sigma',k_y}(E)$ is a transmission probability 
of an electron 
from the spin state $\sigma'$ ($\sigma'=\up$, $\dn$) in the left lead 
to the spin state $\sigma$ ($\sigma=\up$, $\dn$) in the right lead, 
assuming the conservation of $k_y$ during transmission.
In our simple model as mentioned above, we set $k_y=0$ and skip the summation over $k_y$. 
Then, $T_{\up\dn,0}(E)=T_{\dn\up,0}(E)=0$ if we choose the spin basis in $\hat{y}$, because 
$\hat{H}_{k_y}$ commutes with  $\sigma_y$. 
Thus, the spin-up and spin-down transmissions are treated separately.  
The spin-polarization of the electric current 
in the right lead is given by
\begin{eqnarray}
P=\frac{I_\up-I_\dn}{I_\up+I_\dn}=\frac{G_\up-G_\dn}{G_\up+G_\dn}. 
\end{eqnarray}
The actual numerical values of 
$T_{\sigma\sigma',k_y}$ can be calculated 
using the Recursive Green Function 
technique from the TB Hamiltonian [Eq.~(\ref{eq:Hamiltonian2})] 
as ellaborated in Ref.~\onlinecite{nikolic}. 

%

\subsection{Values of the tight-binding parameters and the 
confirmation of the band dispersion of DQW}
The actual material system of our choice to make the proposed device 
is  In$_{0.53}$Ga$_{0.47}$As/In$_{0.52}$Al$_{0.48}$As 
DQW,\cite{koga_IEICE} where the well width $d_{\rm QW}$ for both QW1 
and QW2 is 10 nm and the barrier thickness $d_{\rm B}$ ranges 
1.5 -- 5.0 nm. The donor doping (Si$^+$) in the carrier supplying 
layers above and below the DQW results in the 
electric fields along $z$ axis within the DQW as explained in Sec.~II.
Other material specific parameters such as the energy band gap and 
the conduction band offset are given elsewhere.\cite{Koga} 

The following values are used in our actual calculations. 
$t_{\rm O}=\hbar^2/2m^*_{||}a^2=0.81$ eV from 
$m^*_{||}=0.047 m_e$ and $a=1$ nm. 
$t_{\rm SO}=\alpha_{\rm R}/2a=1.57$ meV, where 
$\alpha_{\rm R}=
\frac{a_{\rm SO}
\left | e \right |N_{\rm S}^{\rm tot}/2}{2\epsilon_{\rm 
S}\epsilon_0}=3.14\times 10^{-12}$ eVm 
assuming $\partial V(z)/\partial z = 0$ 
at the center of the barrier layer between QW1 and QW2 ($z=0$). 
We note $a_{\rm SO}$ (intrinsic constant for the Rashba effect) 
$=25.28$ \AA$^2$ and $\epsilon_{\rm S}$ (dielectric constant) 
$=13.1$ for In$_{0.53}$Ga$_{0.47}$As\cite{note_aSO}. 
The total sheet carrier density $N_{\rm S}^{\rm tot}$ was assumed 
to be 3.6$\times 10^{16}$ m$^{-2}$ throughout the article. 
The values of $t_{\rm coup}$ (a half of the subband splitting) and 
$\left \langle z  \right \rangle\equiv 
\left<\Phi_{\rm QW2}\right|z\left|\Phi_{\rm QW2}\right>$ 
are obtained from the 
self-consistent solutions of the one-band 
Poisson-Schr\"{o}dinger equations of the unperturbed 
Hamiltonian $H_0$ with $B=0$ 
[Eq.~(\ref{eq:hamiltonian_bfield})] assuming the Neumann's boundary 
condition $\partial V(z)/\partial z = 0$ at $z=\pm \left ( 
d_{\rm B}/2+22 \right )$ nm, which are just outside the 6 nm 
thick, symmetrically placed carrier supplying layers above 
and below the DQW, 
where the donor density is $1.8\times 10^{16}$ m$^{-2}$ 
(a half of $N_{\rm S}^{\rm tot}$). 
It turned out that the values of $t_{\rm coup}$ and 
$\left \langle z \right \rangle$  approximately obey 
the following phenomenological equations: 
$t_{\rm coup} ({\rm eV}) =11.66\,e^{-1.09\,d_{\rm B}}$ 
and
$\left \langle z \right \rangle ({\rm nm})
=-1.105\,e^{-1.12\,d_{\rm B}}+5.744+d_{\rm B}/2$, where the unit for 
$d_{\rm B}$ is nm.

\begin{figure}[h]
\hspace*{-2cm}
\includegraphics[width=12.5cm]{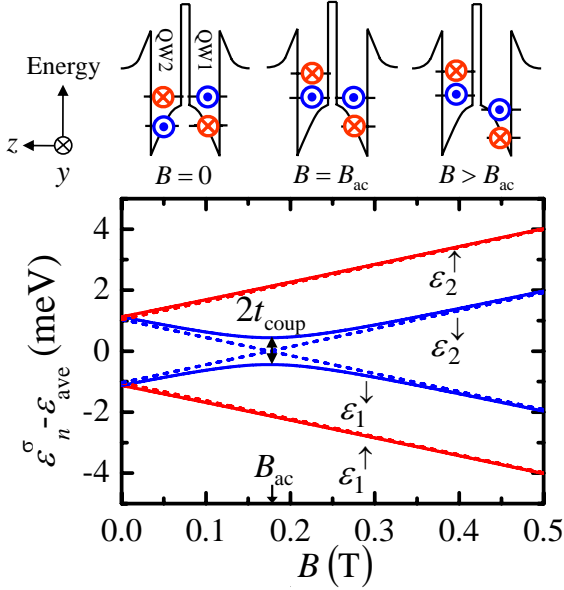}
\caption{(Color online) 
Energy eigenvalues in the symmetric DQW with 
$d_{\rm B}=3$ nm for a given wave vector $\vect{k}=(k_{\rm F},0)$ 
($k_{\rm F}=3.36\times 10^{8}$ m$^{-1}$) as a function of the 
in-plane magnetic field, where the solid and dashed curves 
correspond to the calculations with and without the inter-well coupling 
$t_{\rm coup}$, 
respectively.  
$\ep_{\rm ave}$ stands for the 
averaged energy $\ep_{\rm ave}\equiv (\ep_1^\up+\ep_1^\dn+\ep_2^\up
+\ep_2^\dn)/4$. 
On top of the main figure, we depict the spin-dependent energy levels 
for $t_{\rm coup}=0$
at three representative magnetic fields.  
}
\label{spin_splitting}
\end{figure}
The correspondence between the continuous effective 
mass model [Eq.~(\ref{eq:hamiltonian_bfield})] and the TB model 
[Eq.~(\ref{eq:Hamiltonian2})] {{is made clear}} 
if one diagonalizes 
the following 4$\times$4 TB Hamiltonian of the DQW 
part of the device 
in the 
spin$\otimes$QW space for a unit cell, taking into account 
the $k_x$-dependent Bloch phase assuming the translational 
symmetry along the $x$ direction, which is equivalent to applying 
the periodic boundary condition.
\begin{equation}
H^{{\rm spin}\otimes{\rm QW}}\equiv e^{-ik_xa}H_{0,-1}+H_{0,0}
+e^{ik_xa}H_{0,+1},
\label{Hamiltonian_bulkTB}
\end{equation}
where
\begin{equation}
H_{0,0}=\left (
\begin{array}{cccc}
2t_{\rm O}&0&-t_{\rm coup}&0\\
0&2t_{\rm O}&0&-t_{\rm coup}\\
-t_{\rm coup}&0&2t_{\rm O}&0\\
0&-t_{\rm coup}&0&2t_{\rm O}\\
\end{array}
\right )
\end{equation}
and
\[\hspace{-1.5cm}
\left \langle \begin{array}{c}m_x\\1\up\\\end{array}\right | 
\left \langle \begin{array}{c}m_x\\1\dn\\\end{array}\right | 
\left \langle \begin{array}{c}m_x\\2\up\\\end{array}\right | 
\left \langle \begin{array}{c}m_x\\2\dn\\\end{array}\right | 
\]\hspace{-2cm}
\vspace{-0.5cm}
\begin{equation}
H_{0,\pm1}=\left (
\begin{array}{cccc}
t^{\up\up}_{\rm QW1}
&t^{\up\dn}_{\rm QW1}&0&0\\
t^{\dn\up}_{\rm QW1}&
t^{\dn\dn}_{\rm QW1}&0&0\\
0&0&t^{\up\up}_{\rm QW2}&t^{\up\dn}_{\rm QW2}\\
0&0&t^{\dn\up}_{\rm QW2}&t^{\dn\dn}_{\rm QW2}\\
\end{array}
\right )
\begin{array}{c}
\left | m_x\pm1,m_z=1,\up \right \rangle\\
\left | m_x\pm1,m_z=1,\dn \right \rangle\\
\left | m_x\pm1,m_z=2,\up \right \rangle\\
\left | m_x\pm1,m_z=2,\dn \right \rangle\\
\end{array}\label{eq:4by4}
\end{equation}
are the couplings among the 4 sites in the 
spin$\otimes$QW space at the $m_x$th lattice site 
and those between the $m_x$th and 
$\left ( m_x\pm1 \right )$th lattice 
sites, respectively. In Eq.~(\ref{eq:4by4}), 
the matrix elements $t^{\sigma\sigma'}_{\rm QW1}$ 
and $t^{\sigma\sigma'}_{\rm QW2}$ are
given by 
$t^{\sigma\sigma'}_{\vect{m},\vect{m}\pm\vect{e}_x}$
of Eq.~(\ref{eq:hopping})
with $m_z=1$ and 2 for QW1 and QW2, respectively.

Choosing the spin basis in $\hat{y}$, $H^{{\rm spin}\otimes{\rm QW}}$ 
can be block diagonalized and solved separately for each spin. 
The sorted spin-up and -down Hamiltonians are
\begin{equation}
\begin{array}{l}
H_{\up,\dn}=2t_{\rm O}I_{\rm QW}-2\sqrt{t_{\rm O}^2+t_{\rm SO}^2}\\
\\
\hspace{0.5cm}\times\left (
\begin{array}{cc}
{\rm cos}( k_xa+\tilde{\phi}_{\up,\dn}
) & 
\tilde{t}_{\rm coup} \\
\tilde{t}_{\rm coup}&
{\rm cos}( k_xa-\tilde{\phi}_{\up,\dn}
)
\end{array}
\right )\label{eq:block_diagonalized_hamil}
\end{array}
\end{equation}
in the QW space. We note 
$\tilde{t}_{\rm coup}= 
t_{\rm coup}/2\sqrt{t_{\rm O}^2+t_{\rm SO}^2}$ 
and $\tilde{\phi}_{\up,\dn}= \phi\pm\tilde{\alpha}$ 
($+$ for $\up$ and $-$ for $\dn$), where
$\phi= eBa\left \langle z \right \rangle/\hbar$ 
and
$\tilde{\alpha} = {\rm tan}^{-1}\left ( t_{\rm SO} / t_{\rm O} \right )$.
$I_{\rm QW}$ is the $2 \times 2$ identity matrix in the QW space.
The eigenvalues of this {{Hamiltonian}}, 
which provide the spin-dependent energy dispersion 
relations for the DQW system in the presence of the Rashba SOI and 
in-plane magnetic field, are
\begin{equation}
\begin{array}{l}
\ep_{1,2}^{\up,\dn} 
=2t_{\rm O}-2\sqrt{t_{\rm O}^2+t_{\rm SO}^2}{\cos}k_xa\,{\rm cos}\tilde{\phi}_{\up,\dn}\\
\\
\hspace{0.5cm}\pm \sqrt{t_{\rm coup}^2 + 4
\left ( t_{\rm O}^2 + t_{\rm SO}^2 \right )
\left (
{\rm sin}k_xa\,{\rm sin}\tilde{\phi}_{\up,\dn}
\right )^2},
\end{array}
\label{eq;analytical_energy}
\end{equation}
where $-$ and $+$ signs are for subbands 1 and 2, respectively.
One finds that these energies agree with the results 
of the effective mass model in 
Eq.~(\ref{eq:hamiltonian_bfield}) within 1\% at the Fermi energy by
Taylor expanding the trigonometric functions 
in Eq.~(\ref{eq;analytical_energy}), 
where the following values are sufficiently smaller than unity: 
$k_xa\approx k_{\rm F}a = 0.336$, $\phi \approx {4.3}\times 10^{-3}$ 
(for {{$B=0.2$}} T and $\left \langle z \right \rangle 
= 7$ nm) and 
$\tilde{\alpha}\approx t_{\rm SO}/t_{\rm O} = 1.9 \times 10^{-3}$.

In Fig.~\ref{spin_splitting}, we plot the magnetic field 
dependence of these eigenenergies for $d_{\rm B}=3$ nm 
($t_{\rm coup}=0.44$ meV) and $k_x=k_{\rm F}=3.36\times10^8$ m$^{-1}$ 
relative to their averaged value, together with the results when the 
coupling between QW1 and QW2 is turned off ($t_{\rm coup}=0$). 
Here energy difference between the spin degenerate 
pairs $\ep_2^{\up,\;\dn}$ and $\ep_1^{\up,\:\dn}$ at $B=0$, which is 
interpreted as the subband splitting in the symmetric DQW, is very 
close in value to the Rashba splitting energy 
2$\alpha_{\rm R}k_{\rm F}$ of independent QW1 and QW2. 
This is because the orbital parts of the eigenfunctions for each spin-degenerate 
subband level are localized in either QWs depending on their spin state 
[Fig.\ref{fig_wfunc}(a)]. 
The slopes for $B\rightarrow0$~T 
in Fig.~\ref{spin_splitting} 
with $t_{\rm coup}=0$ are given exactly by 
$\pm\hbar e\left<z\right>k_{\rm F}/m^*_{||}= \pm 5.97$ meV/T. }
Upon increasing $B$, Fig.~\ref{spin_splitting} shows that the energy levels 
$\ep_1^\dn$ and $\ep_2^\dn$ anticross each other at  
$B=B_{\rm ac}\equiv m_{||}^*\alpha_{\rm R}/\hbar e\left<z\right>=0.177$ T, 
where $\tilde{\phi}_\dn=\phi-\tilde{\alpha}=0$ and 
Eq.~(\ref{eq;analytical_energy}) reduces to 
$\ep_{1,2}^{\dn}=2t_{\rm O}-2\sqrt{t_{\rm O}^2+t_{\rm SO}^2}{\cos}k_xa
\pm t_{\rm coup}$.  
This means that the anticrossing energy gap 
found in Fig.~\ref{spin_splitting} is exactly the subband splitting energy 
in the absence of both the Rashba SOI 
and the magnetic field [Fig.~\ref{fig_wfunc}(a)]. 
Thus the bonding and antibonding states 
are formed selectively for spin-down electrons. 
Spin-up wavefunctions are then localized in either QW1 or QW2. 
As we further increase the magnetic field ($B>B_{\rm ac}$), 
both the spin-up and $\mbox{-down}$ wave functions are localized in either QW1 or QW2 
(see the top pictures in Fig.~\ref{spin_splitting}).

\section{Transport analysis and Spin filtering properties}
%
\begin{figure}[h]
\hspace*{-1.2cm}
\includegraphics[width=15cm]{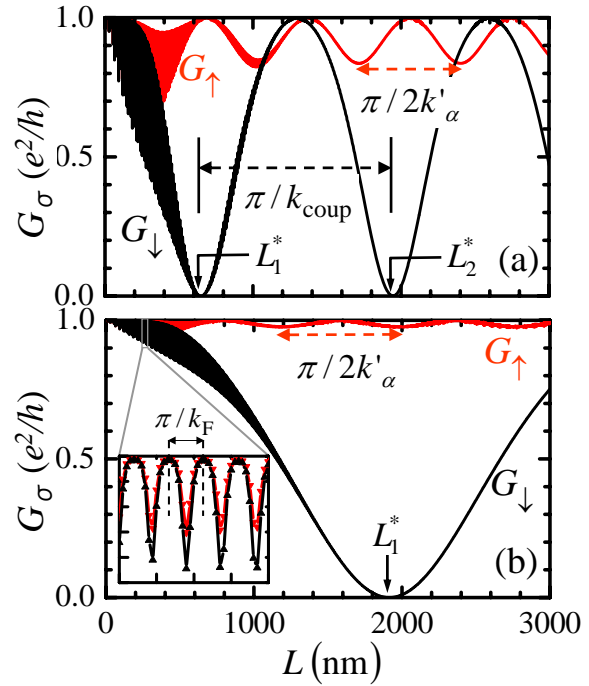}
\caption{(Color online) Plots of the 
spin-dependent conductances $G_{\up,\dn}$, calculated for 
two different barrier thicknesses (a) 
$d_{\rm{B}}=2$ nm and (b) $d_{\rm{B}}=3$ nm as a function of the device length $L$.
The magnetic fields are 
fixed at their anticrossing values 
{{$B_{\rm ac}=0.192$ T for (a) and 0.177 T for (b)}. 
The inset of (b) is the magnified view of the main panel between 
$260$ nm and  $300$ nm for abscissa and between 
$0.9~e^2/h$ and $1.0~e^2/h$ for ordinate, 
where the period of the 
rapid oscillations is found to be $\pi/k_{\rm F}$.}
}
\label{conductance}
\end{figure}
In Fig.~\ref{conductance}, we plotted the calculated values of 
the spin-dependent conductances $G_{\up,\dn}$ ($\up\parallel\hat{y}$) 
as a function of $L$ at $B=B_{\rm ac}$ for $d_{\rm{B}}=2$ and 
3 nm ($T=$5 K). 
We find that the change in the value of $G_\dn$ 
as a function of $L$ is more pronounced than that of $G_\up$, 
where $G_\dn$ even becomes zero at specific device lengths $L^*_n$, 
$n$ being an integer, while $G_\up$ varies only weakly 
with $L$. These qualitative behaviors are in agreement with 
the argument in Sec.~II. 
We find that $L^*_1 = 648$~nm and $L^*_2=1946$~nm for 
$d_{\rm B}=2$ nm, whereas $L_1=646$~nm and $L_2=1939$~nm in Sec.~III. 
Similarly, $L^*_1=$ 1914 nm here agrees with 
$L_1=1906$ nm in Sec.~II for $d_{\rm B}=3$ nm.  
We note $L_n\equiv\pi\left (n-\frac{1}{2} 
\right )/k_{\rm coup} $, where 
$k_{\rm coup}=t_{\rm coup}(dE(k)/dk)^{-1}
\approx t_{\rm coup}m^*_{||}a/\hbar^2\sin(k_{\rm F}a)$
in the TB model.

\begin{figure}[!h]
\hspace*{-1.5cm}
\includegraphics[width=15cm]{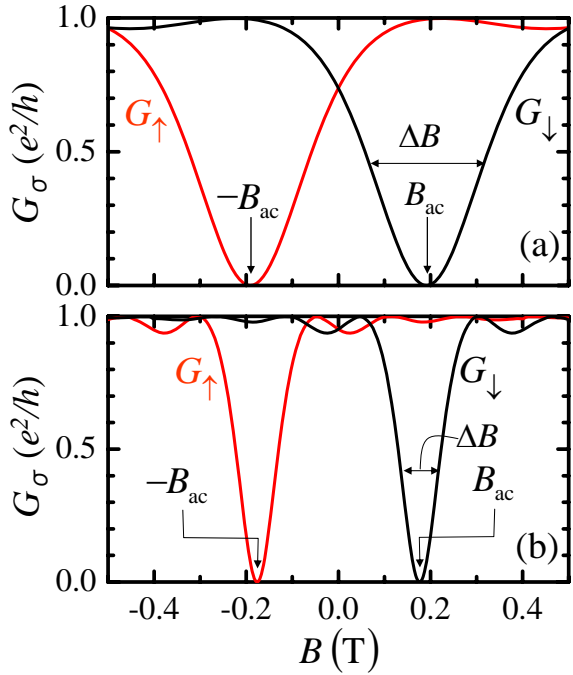}
\caption{
(Color online) 
Magnetic field dependence of the spin-dependent conductances 
for  (a) $d_{\rm{B}}=2$ nm and (b) $d_{\rm{B}}=3$ nm.
The device lengths for $d_{\rm B}=2$ nm and 3 nm are 
{{$L=646$ nm and 1906 nm}}, respectively. 
The anti-crossing magnetic fields $B_{\rm ac}$ 
for $d_{\rm B}=2$ nm and 3~nm are 
$0.192$ T and 0.177 T, respectively, 
as indicated in the figures. 
See text for the explanation of $\Delta B$. 
}
\label{bdependence}
\end{figure}

Shown in Fig.~\ref{bdependence} are the magnetic field dependence of 
$G_{\up,\dn}$ for  $d_{\rm{B}}=2$ nm and 3 nm, where the values of 
$L$ are 646 and  1906 nm, respectively. 
For $B>0$, we observe that $G_{\dn}$ 
becomes zero at $B=0.192$ T and 0.177 T 
for $d_{\rm B}=2$ nm and 3 nm, respectively, i.e., at the 
anti-crossing magnetic field $B_{\rm ac}$.
For $B<0$, on the other hand, $G_{\up}$ reaches zero at 
$B=-B_{\rm ac}$ while 
$G_{\dn}$ is kept close to unity, { as is consistent with the 
symmetry found in the Hamiltonian [Eq.~(\ref{eq:hamiltonian_bfield})}]. 
In Fig.~\ref{bdependence}, we also recognize that 
the variation of $G_{\dn}$ ($G_{\up}$) around 
$B=B_{\rm ac}$ ($B=-B_{\rm ac}$) 
is more moderate for $d_{\rm B}=2$ nm than for 
$d_{\rm B}=3$ nm.  This is because the bonding/antibonding states 
($1/\sqrt{2} \left \{ \left | \Psi_b \right \rangle 
\pm \left | \Psi_a \right \rangle \right \}$ ) at 
$B=\pm B_{\rm ac}$ are more robust with the change of $B$
for $d_{\rm B}=2$ nm than for $d_{\rm B}=3$ nm 
because of the stronger inter-well coupling $t_{\rm coup}$.
The range of the magnetic field 
$\Delta B$ around $B=B_{\rm ac}$ within which the bonding/antibonding states 
persist can be roughly estimated by equating 
$e\Delta B \left \langle z \right \rangle/\hbar$ to $k_{\rm coup}$. 
We obtain $\Delta B=0.24$ T and 0.075 T for 
$d_{\rm B}=2$ nm and $d_{\rm B}=3$ nm, respectively, which agree well with 
the widths of conductance dips observed in Fig.~\ref{bdependence}.  
Thus, measuring the width of the magneto-conductance dip 
would provide an estimate of the value of $t_{\rm coup}$.

We note that there are features 
in Fig.~\ref{conductance} that cannot be explained by the simple 
flying qubit model (Sec.~II). 
The first is the rapid oscillation of the spin-dependent 
conductance as shown in the inset of Fig.~\ref{conductance}(b).
We notice that these oscillations (i) are more pronounced for 
shorter $L$, (ii) disappear as $G_{\up,\dn}$ approaches to
zero or the maximum value $e^2/h$, and (iii) damp away with increasing $L$.
The period of the rapid oscillation, which is essentially 
constant with $L$, is found to be $\Delta L\simeq 9.34$ nm.\cite{rapid_oscillation_period} 
This is in good agreement with the value of $\pi/k_{\rm F}$, 
which infers some resonance phenomena associated with the plane wave part of the 
electron wave function. 
Detailed analysis using the multiple reflection model (Appendix B) 
revealed that it is not the 
discontinuities in the values of $B$ and/or $\alpha_{\rm R}$ 
in the Hamiltonian between the leads and the DQW 
but the finite probability amplitude of the elctron wave function 
within QW2 which resulted from tunneling from QW1,  
that caused the wave function reflection at $x=0$ and $L$. 
The reason for the disappearance of the 
rapid oscillations as $G_{\dn}$ approaches to zero or $e^2/h$
is the disappearance of the wave 
function amplitude within QW2 at either $x=0$ or $L$ 
with the condition $L=L^*_n$ or $L=L^*_n+\pi/2k_{\rm coup}$, respectively.
\cite{note_Gup_oscillation}

Second, the damping of the rapid oscillations with increasing $L$ 
in Fig.~\ref{conductance} is {due} to 
the finite temperature assumed in our calculation. 
Considering the finite width $\Delta k$ in the wave 
number values that participates in the electron transport, 
the criteria {for having} this rapid oscillation 
is $\Delta k L \ll 1$. Using 
$\Delta k = k_{\rm B}T\left ( \partial E/\partial k \right ) ^{-1} 
\approx 8\times 10^{-5}$ m$^{-1}$ for $T=5$ K, we obtain 
$L\ll 1.25$ $\mu$m, which explains the decay of the rapid 
oscillation with increasing $L$.

Another feature in Fig.~\ref{conductance} that cannot be explained by 
the simple flying qubit model in Sec.~II is the weak 
modulation of the spin-up conductance $G_\up$ with period $\pi/2k'_\alpha$ in  
Fig.~\ref{conductance}, 
where $k_\alpha'$ is the corrected Rashba wave number (see below). 
This modulation of $G_\up$ can be explained qualitatively 
by the following extended qubit model 
{ (see Appendeix B for the exact model).}
It turned out that even the eigenfunctions of the spin-up electrons are more correctly 
described as superpositioned states between QW1 and QW2 because the inter-well 
coupling $t_{\rm coup}$  is not completely negligible for the spin-up electrons either.
These eigenfunctions still keep  the features of 
antibonding-like or bonding-like wave functions in a sense whether they 
have a node or not, respectively, 
\begin{eqnarray}
\left|\Phi_{\rm a}^\up\right>
&=&\delta \left|\Phi_{\rm QW1}^\up\right>
-
\sqrt{1-\delta^2}\left|\Phi_{\rm QW2}^\up\right>, \\
\left|\Phi_{\rm b}^\up\right>
&=&
\sqrt{1-\delta^2}\left|\Phi_{\rm QW1}^\up\right>
+\delta\left|\Phi_{\rm QW2}^\up\right>,
\end{eqnarray}
where $0<\delta< 1$, apart from the plane wave part for the in-plane transport. 
We obtained {{$\delta=0.275$ and 0.104}} for $d_{\rm B}=2$ nm and 3 nm, 
respectively, from the eigenvectors of the spin-up Hamiltonian 
[Eq.~(\ref{Hamiltonian_bulkTB})].
For electrons propagating in the positive $x$ direction, the in-plane wave 
numbers for $|\Phi_{\rm b}^\up\rangle$ and 
$|\Phi_{\rm a{}}^\up\rangle$ are 
$k_{\rm F}+2k_\alpha'$ and $k_{\rm F}-2k_\alpha'$, respectively, 
where the corrected Rashba wave number $k_\alpha'$ is 
$k_\alpha+k_{\rm coup}^2/8k_\alpha$ incorporating the 
contribution of the inter-well coupling $t_{\rm coup}$ to the wave 
number shift (see Appendix A).  
After similar procedures as 
in Sec.~II, which we call as the 
extended flying qubit model, 
we obtain the following expression for the 
transmission probability of the spin-up electrons, 
\begin{equation}
T_{\up(\rm QW1)}=1-
4\delta^2(1-\delta^2)\sin^2\left (2k'_\alpha L \right ).   
\label{eq:Transmission_dn2}
\end{equation}
Thus, the periods of conductance modulation for spin-up electron are 
$\pi/2k'_\alpha=677$ nm and 792 nm for 
$d_{\rm B}=2$ nm and 3 nm, respectively, which are in good agreement with
683 nm and 799 nm in the TB calculation (Fig.~\ref{conductance}). 
We note that the extended flying qubit model overestimates the oscillation 
amplitudes of $G_\up$, i.e., $4\delta^2(1-\delta^2)=0.2805$ and 0.0429 
in units of $e^2/h$ for $d_{\rm B}=2$ nm and 3 nm, respectively, whereas 
the corresponding values in the TB calculations {are} about 
half of these. The discrepancy comes from the absence of both 
the multiple reflections of wave function between $x=0$ and $L$ 
and 
the thermal averaging effect in our naive model. 
We derived the exact results assuming multiple reflections 
between $x=0$ and $L$ {which reproduced the TB results} in Appendix B.

\section{Conclusion}
We proposed a lateral spin-blockade device using InGaAs/InAlAs double quantum well (DQW), 
where the values of the Rashba spin-orbit parameter $\alpha_{\rm R}$ are opposite in 
sign but equal in magnitude between the constituent quantum wells (QW). 
The principle of the spin-blocking effect in the proposed device is in 
the spin-selective matching of the front and back edges of the spin-split 
Fermi circles [Fermi circle points at $(\pm k_{\rm F}, 0)$] between the two QWs, 
which is made possible by the in-plane magnetic field ${\bf B}=(0, B_{\rm ac}, 0)$. 
The superposition of the resulting bonding and antibonding wavefunctions 
that are formed for the selected (e.g., spin-down) electrons exclusively results in the 
precessional motion of electrons between the QW1 and QW2, which is denoted 
as the ``flying qubit" state. The ``flying qubit" state can be blocked by depleting 
or etching away only the QW2 part at a length of half-integer multiple of 
the precession wave length as depicted in Fig.~\ref{fig_trajectory_bfield}(b). 

We would like to reiterate the features of the proposed spin-filtering device. 
(i) The proposed device is novel in a sense that it operates based on the conventional 
band theory and the Boltzmann transport theory, adding the spin degree of freedom in the 
form of the Rashba effect. Thus, only the elementary level of quantum mechanics and 
solid state physics is required to understand the principle of the device. 
(ii) In our simplified one-dimensional model where two singly-channeled leads are attached to the device, 
we obtain a perfect spin-blockade by the in-plane magnetic field ${\bf B}=(0, B_{\rm ac}, 0)$, 
whereas such spin-blocking fails to happen in the case of electrical control using the gate. 
The latter observation is found to be consistent with the recent theoretical 
results proven analytically for devices with two singly-channeled leads. 
(iii) The actual devices, however, 
will be prepared in the form of the two-dimensional electron gas (2DEG), 
where finite $k_y$ components ($\perp {\bf I}$) in the electron wave vector also 
participate in the electron transport. 
While some deterioration in the spin-polarization will be expected in the actual 
2D model due to the finite $k_y$ components, the multi-channeled nature of the device would, 
in turn, allow the generation of spin-polarized current purely electrically. 
We can expect that the spin-polarized currents thus generated are fairly large due 
to the multi-channeled nature of the device, as compared to those generated by the QPC-based devices, 
for example, which open the possibilities of future spintronics applications widely.

\section*{Acknowledgment}

This work was supported by KAKENHI, Grant-in-Aid for Scientific Research (B), Grant No. 23360001.

{

\section*{Appenxix A : 
{Correction to the  Rashba wave number $k_\alpha$
due to the inter-well coupling $t_{\rm coup}$}}
The generalized Rashba wave number 
$k_\alpha^\prime$ is associated with the energy 
difference $\ep_{2}^{\up} -\ep_{1}^{\up} $ 
with $\vect{B}=\left (0, B_{\rm ac},0\right )$
at the Fermi wave number as following [see Eq.~(\ref{eq;analytical_energy})].
\begin{eqnarray}
\ep_{2}^{\up} 
-
\ep_{1}^{\up} 
&=&
2\sqrt{t_{\rm coup}^2 + 4
\left ( t_{\rm O}^2 + t_{\rm SO}^2 \right )
\left(\sin k_{\rm F}a \sin \tilde{\phi}_\dn\right)^2}
\nonumber \\
&\equiv&
4k_\alpha'
\left.\frac{\partial E(k_x)}{\partial k_x}\right|_{k_x=k_{\rm F}}
\label{eq_appAeq1}
\end{eqnarray}
In the limit $a\to 0$, we have
$\tilde{\phi}_\up=\phi+{\rm tan}^{-1}\left ( t_{\rm SO}/t_{\rm O} \right ) 
= 2t_{\rm SO}/t_{\rm O}$, where $\phi=eB_{\rm ac}a\left \langle z \right 
\rangle/\hbar = t_{\rm SO}/t_{\rm O}$, $t_{\rm SO}/t_{\rm O}=k_\alpha a$, 
and $t_{\rm coup}=\hbar^2k_{\rm F}k_{\rm coup}/m^*_{||}$. Note
$t_{\rm O}=\hbar^2/ 2m^*_{||}a^2$. Substituting these in  
Eq.~(\ref{eq_appAeq1}) and letting  $a\to 0$, we have 
\begin{eqnarray}
\ep_{2}^{\up} 
-
\ep_{1}^{\up} 
&=&
\frac{4\hbar^2k_{\rm F}xk_\alpha}{m^*_{||}}
\sqrt{1+\frac{k_{\rm coup}^2}{4k_\alpha^2}}
\nonumber \\
&\approx&
4k_\alpha \left ( 1
+\frac{k_{\rm coup}^2}{8k_\alpha^2} \right )
\left . \frac{\partial E(k_x)}{\partial k_x}\right|_{k_x=k_{\rm F}}, 
\label{eq_appAeq2}
\end{eqnarray}
for $ k_{\rm coup} < k_\alpha$. 
Comparing Eqs.~(\ref{eq_appAeq1}) and (\ref{eq_appAeq2}), we obtain
\begin{eqnarray}
k'_\alpha
\approx
k_\alpha
+
\frac{k_{\rm coup}^2}{8k_\alpha}. 
\end{eqnarray}
}

\section*{Appendix B: Multiple reflection model}
Consier a DQW device with length $L$. Let $t_{mm^\prime}$ ($m,m^\prime=1,2$) 
be the nominal quantum mechanical transmission amplitude of electron 
from $x=0$ in QW$m^\prime$ to $x=L$ in QW$m$. Similarly we let 
$r_{mm^\prime}$ be the nominal transmission amplitude of electron 
from $x=L$ in QW$m^\prime$ to $x=0$ in QW$m$.
The multiple reflection model states that the overall 
trasmission amplitude from QW1 at $x=0$ to QW1 at $x=L$ is given by
\begin{eqnarray}
t_{\rm tot}&=& t_{11}+t_{12}r_{22}t_{21}+t_{12}r_{22}t_{22}r_{22}t_{21}+\cdot\cdot\cdot
\nonumber \\
&=&t_{11}+t_{12}r_{22}\left \{ 1+ \left ( t_{22}r_{22} \right ) +
\left ( t_{22}r_{22} \right )^2+\cdot\cdot\cdot \right \}t_{21}
\nonumber  \\
&=&t_{11}+\frac{t_{12}r_{22}t_{21}}{1- t_{22}r_{22}}, 
\label{eq_ttot}
\end{eqnarray}
where the transmission probability $T_{\rm tot}$ is given by $\left | t_{\rm tot} \right |^2$. 
The values of $t_{mm^\prime}$ and $r_{mm^\prime}$ per spin are obtained 
by the generalized flying qubit model as following.

Choosing the spin quantization axis in $\hat{y}$, the eigenfunctions in DQW 
without spin 
are  $e^{\pm i \left ( k_{\rm F} + \Delta k \right ) 
x}\left|\Phi_{\rm b}\right\rangle$
and
$e^{\pm i \left ( k_{\rm F} - \Delta k \right ) 
x}\left|\Phi_{\rm a}\right\rangle$
at $E=E_{\rm F}$ and $\vect{B}=\left ( 0, B_{\rm ac}, 
0\right )$, where
\begin{eqnarray}
\left|\Phi_{\rm b}\right\rangle
&=&
\sqrt{1-\delta^2}
\left|\Phi_{\rm QW1}\right\rangle
+
\delta
\left|\Phi_{\rm QW2}\right\rangle,
\\
\left|\Phi_{\rm a}\right\rangle
&=&
\delta
\left|\Phi_{\rm QW1}\right\rangle
-
\sqrt{1-\delta^2}
\left|\Phi_{\rm QW2}\right\rangle
\end{eqnarray}
($0 < \delta < 1 $). We note $\Delta k = k_{\rm coup}$ or $2k_\alpha^\prime$ 
for spin-up or spin-down electrons, respectively, and 
${}^t\hspace{-0.1cm}\left ( \sqrt{1-\delta^2}, \delta \right )$
and
${}^t\hspace{-0.1cm}\left ( \delta, -\sqrt{1-\delta^2} \right )$
are the eigenvectors of Eq.~(\ref{eq:block_diagonalized_hamil}). 
We solve these for 
$\left|\Phi_{\rm QW1}\right\rangle$ and 
$\left|\Phi_{\rm QW2}\right\rangle$.
\begin{eqnarray}
\left|\Phi_{\rm QW1}\right\rangle
&=&
\sqrt{1-\delta^2}\left|\Phi_{\rm b}\right\rangle
+
\delta
\left|\Phi_{\rm a}\right\rangle \\
\left|\Phi_{\rm QW2}\right\rangle
&=&
\delta \left|\Phi_{\rm b}\right\rangle
-
\sqrt{1-\delta^2}\left|\Phi_{\rm a}\right\rangle 
\end{eqnarray}
We let these wave functions propagete from $x=0$ to $x=L$.
\begin{eqnarray}
&&\left|\Phi_{\rm QW1}\right\rangle\;({\rm at}\;x=0)
\nonumber \\
&&\rightarrow
e^{ik_{\rm F} L}\left \{e ^{i\Delta k L}
\sqrt{1-\delta^2}\left|\Phi_{\rm b}\right\rangle
+
e^{-i\Delta k L}\delta
\left|\Phi_{\rm a}\right\rangle \right \}
\nonumber \\
&&
=e^{ik_{\rm F} L}\left [ \left \{
e ^{i\Delta k L} 
-2i\delta^2{\rm sin}\left ( {\Delta k L} \right )\right \} 
\left|\Phi_{\rm QW1}\right\rangle \right .
\nonumber \\
&&
\hspace{2cm}\left .+2i\delta\sqrt{1-\delta^2}{\rm sin}\left ( {\Delta k L} \right )
\left|\Phi_{\rm QW2}\right\rangle \right ]
\nonumber \\
&&
\hspace{1.3cm}({\rm at}\;x=L)
\nonumber \\
\nonumber \\
&&
\left|\Phi_{\rm QW2}\right\rangle\;({\rm at}\;x=0)
\nonumber \\
&&\rightarrow
e^{ik_{\rm F} L}\left \{e ^{i\Delta k L}
\delta\left|\Phi_{\rm b}\right\rangle+
e^{-i\Delta k L}\sqrt{1-\delta^2}
\left|\Phi_{\rm a}\right\rangle \right \}
\nonumber \\
&&=e^{ik_{\rm F} L}\left [ 
2i\delta\sqrt{1-\delta^2}{\rm sin}\left ( {\Delta k L} \right )
\left|\Phi_{\rm QW1}\right\rangle\right.
\nonumber \\
&&\hspace{1cm}\left.+\left \{ e ^{-i\Delta k L} 
+ 2i\delta^2{\rm sin}\left ( {\Delta k L} \right ) \right \}
\left|\Phi_{\rm QW2}\right\rangle \right ]
\nonumber\\
&&
\hspace{1.3cm}({\rm at}\;x=L)
\label{eq:PrecessMotion}
\end{eqnarray}
From these we obtain
\begin{eqnarray}
t_{11}&=&
e^{ik_{\rm F}L}
\left\{
e^{i\Delta k  L}
-2i\delta^2\sin(\Delta k L)
\right\},
\label{eq_t11}
\\
t_{12}&=&
e^{ik_{\rm F} L}
\left\{
2i\delta \sqrt{1-\delta^2}
\sin(\Delta k L)
\right\}, 
\label{eq_t12}
\\
t_{21}&=&e^{ik_{\rm F}L}
\left \{
2i\delta\sqrt{1-\delta^2}
\sin(\Delta k L)
\right \}
\label{eq_t21}
\\
\;{\rm and}
\nonumber \\
t_{22}&=&
e^{ik_{\rm F} L}
\left \{
e^{-i\Delta k  L}
+
2i\delta^2
\sin(\Delta k L)
\right \}. 
\label{eq_t22}
\end{eqnarray}
Now we let 
$\left|\Phi_{\rm QW2}\right\rangle$ propagate from $x=L$ 
to $x=0$ to obtain $r_{22}$.    
\begin{figure}[t]
\hspace*{-0.5cm}
 \begin{minipage}{0.5\hsize}
  \begin{center}
   \includegraphics[width=4.5cm]{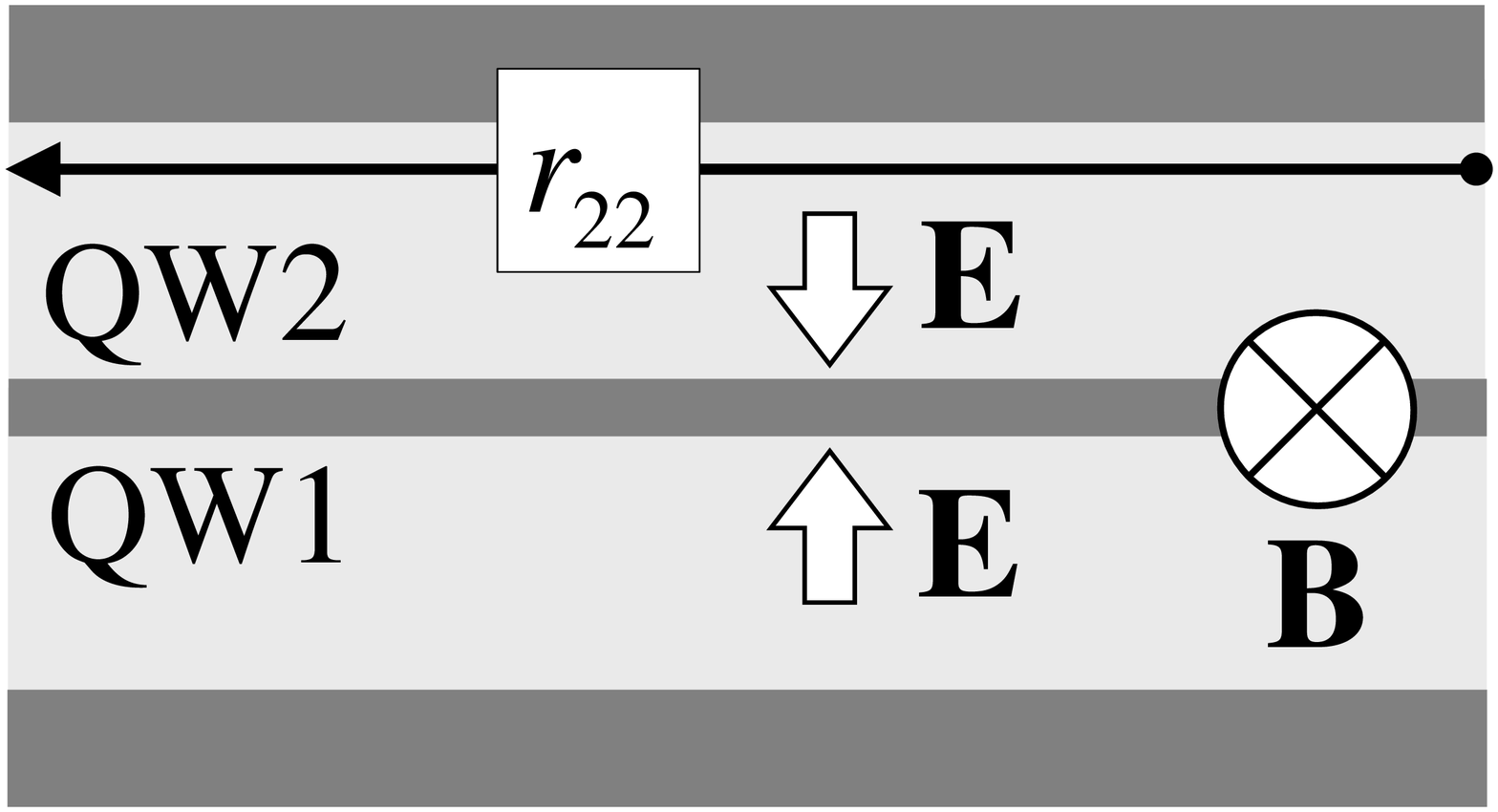}
  \end{center}
 \end{minipage}
\hspace{-0.0cm}
 \begin{minipage}{0.5\hsize}
  \begin{center}
   \includegraphics[width=4.5cm]{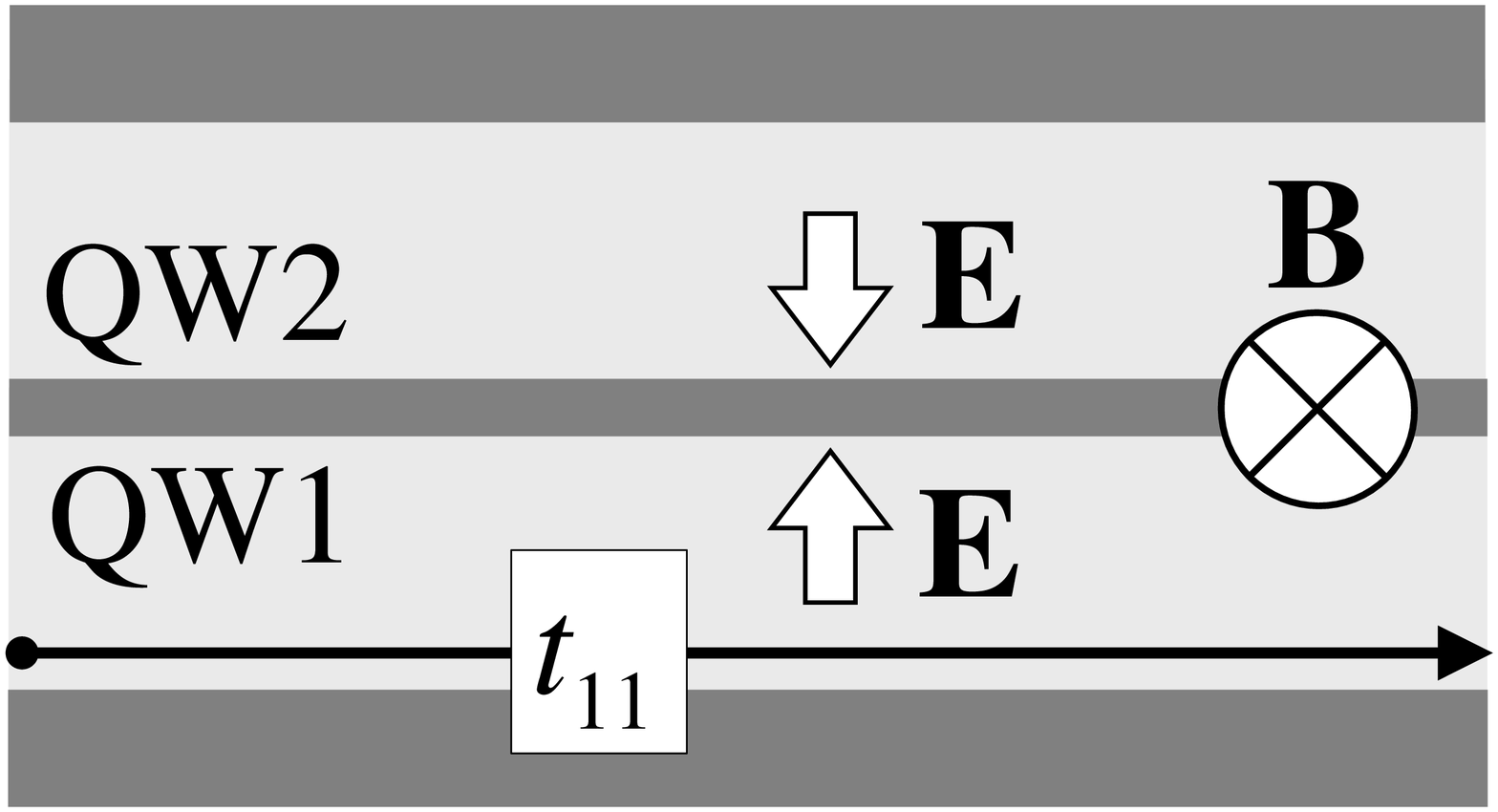}
  \end{center}
  \end{minipage}
\caption{
Schematic illustration for symmetry consideration. 
The nominal transmission amplitude through QW2 with negative $k_x$ ($r_{22}$ in the left panel) 
is found to be equal to that though QW1 with positive $k_x$ ($t_{11}$ in the right panel). }
\label{fig_appB}
\end{figure}
After a simple symmetry consideration in Fig.~\ref{fig_appB}, we notice that it is 
equal to $t_{11}$.  
\begin{eqnarray}
r_{22}&=&t_{11}=
e^{ik_{\rm F} L}
\left \{
e^{i\Delta k  L}
-
2i\delta^2
\sin(\Delta k L)
\right \}. 
\label{eq_r22}
\end{eqnarray}
Substituting Eqs.~(\ref{eq_t11})--(\ref{eq_t22})
and (\ref{eq_r22}) 
into Eq.~(\ref{eq_ttot}), we have
\begin{eqnarray}
t_{\rm tot}=\frac{e^{ik_{\rm F}L}\left(1-e^{2i k_{\rm F} L}\right)
\left[e^{i\Delta k L}-i\delta^2 \sin(k_{\rm F} L)\right]}{1-e^{2ik_{\rm F} L}T_{11}}, 
\label{eq_ttot2}
\end{eqnarray}
where $T_{11}=\left |t_{11}\right |^2
=1-4\delta^2(1-\delta^2)\sin^2(\Delta k L)$ is the result of the extended 
qubit model.  
We obtain the overall transmission probability as
%
\begin{eqnarray}
T_{\rm tot}=|t_{\rm tot}|^2 &=&
\frac{2\left[1-\cos(2k_{\rm F}L)\right]T_{11}}{1-2T_{11}\cos(2k_{\rm F}L)
+T_{11}^2}. 
\label{eq_transmission_analytical}
\end{eqnarray}
We confirmed that this reproduces the TB results 
quantitatively. 
We note that the spin-dependence of $T_{\rm tot}$ is 
realized only through $T_{11}$. Thermal average of $T_{\rm tot}$ is also 
straightforward.
%

\vspace{0cm}
\bibliographystyle{prb}

\begin{references}
\bibitem{note_efield}
$\left \langle E_z 
\right \rangle_{1,2} \equiv \left \langle \Phi_{\rm QW1,2} \right | 
E_z(z) \left | \Phi_{\rm QW1,2} \right \rangle $, where $\left | 
\Phi_{\rm QW1,2} \right \rangle \equiv 1/\sqrt{2} \left \{ 
\left | \Psi_{b} \right \rangle \pm \left | \Psi_{a} \right \rangle \right 
\}$, letting $\left | \Psi_{b} \right \rangle$ and 
$\left | \Psi_{a} \right \rangle$ be the lowest (bonding-like) and 
the second lowest (antibonding-like) energy eigenstates of the unperturbed 
DQW Hamiltonian 
{{along $z$ direction (normal to the QW plane)}} 
with $B=0$ [see Eq.~(\ref{eq:hamiltonian_bfield})]. The sign in $\left \{ 
\left | \Psi_{b} \right \rangle \pm \left | \Psi_{a} \right \rangle \right 
\}$ should be chosen in such a way that 
$\left | \Phi_{\rm QW1} \right \rangle$ and 
$\left | \Phi_{\rm QW2} \right \rangle$ have values in QW1 and QW2, respectively.
\bibitem{note0}
Assuming a plane wave for the in-plane part of the wave function, 
the operators $\frac{\partial}{i\partial x}$ and 
$\frac{\partial}{i\partial y}$ are replaced with the $c$-numbers 
$k_x$ and $k_y$, respectively.
\bibitem{2Dmodel}
We note that the theorem given in Ref.~\onlinecite{gate} does not 
prohibit the generation of spin-polarized current 
for devices with multi-channeled leads, which leaves possibilities 
of generation of spin-polarized current soley by gate in our DQW device experimentally. 
Research in this direction is underway.
\bibitem{note_aSO}
This value of $a_{\rm SO}$ was obtained by substituting the value  
$N_{\rm S}^{\rm tot}/2$ to $N_{\rm S}$ in the formula 
$a_{\rm SO}m^*_{||}/m_{\rm e}=a-bN_{\rm S}$ with 
$a= 1.46e$\AA$^2$ and $b = 1.51\times 10^{-17}e$\AA$^2$m$^2$ 
(see Ref.~\onlinecite{faniel}), 
where we used $m^*_{||}=0.047 m_{\rm e}$.
\bibitem{rapid_oscillation_period}
The oscillation period presented here is an averaged value of 
more than 100 periods.
\bibitem{note_Gup_oscillation}
Similarly, the disappearance of the rapid oscillation in 
$G_\up$ as it approaches to 
$e^2/h$ is due to the disappearance of the wave 
function amplitude in QW2 at $x=L$ when $L=n\pi/2k'_\alpha$.
\end{references}

\end{document}